\documentclass[11pt]{article}

\usepackage[margin=1in]{geometry}
\usepackage[T1]{fontenc}
\usepackage[utf8]{inputenc}

\usepackage{moreverb}

\usepackage{graphicx} % Required for inserting images
\usepackage{grffile}  % safer handling of figure filenames containing spaces
\usepackage{float} % to make figure stick with the text
\usepackage{tabularx}
\usepackage{booktabs} % \toprule, \midrule, \bottomrule
\usepackage{makecell} % for newline in table-cell
\usepackage{tablefootnote} % for table footnotes
\usepackage{endnotes}
\usepackage{textcomp}  % needed for the use of \textunderscore in references \ref{}

\usepackage{array}
\usepackage{authblk}
\usepackage{todonotes}

\usepackage{natbib}

\usepackage{url}
\usepackage{xurl} % better line-breaking for long URLs
\usepackage[colorlinks,bookmarksopen,bookmarksnumbered,citecolor=red,urlcolor=red]{hyperref}

\usepackage[breakable]{tcolorbox}
\usepackage{xcolor}
\usepackage{fancyvrb}
\usepackage{fvextra} % adds breaklines/breakanywhere keys for Verbatim
\usepackage{upquote}

\DefineVerbatimEnvironment{Highlighting}{Verbatim}{commandchars=\\\{\}}

\definecolor{incolor}{HTML}{303F9F}
\definecolor{outcolor}{HTML}{D84315}
\definecolor{cellborder}{HTML}{CFCFCF}
\definecolor{cellbackground}{HTML}{F7F7F7}

\makeatletter
\newcommand{\boxspacing}{\kern\kvtcb@left@rule\kern\kvtcb@boxsep}
\makeatother
\newcommand{\prompt}[4]{}

\makeatletter
\def\PY@reset{\let\PY@it=\relax \let\PY@bf=\relax%
    \let\PY@ul=\relax \let\PY@tc=\relax%
    \let\PY@bc=\relax \let\PY@ff=\relax}
\def\PY@tok#1{\csname PY@tok@#1\endcsname}
\def\PY@toks#1+{\ifx\relax#1\empty\else%
    \PY@tok{#1}\expandafter\PY@toks\fi}
\def\PY@do#1{\PY@bc{\PY@tc{\PY@ul{%
    \PY@it{\PY@bf{\PY@ff{#1}}}}}}}
\def\PY#1#2{\PY@reset\PY@toks#1+\relax+\PY@do{#2}}

\@namedef{PY@tok@w}{\def\PY@tc##1{\textcolor[rgb]{0.73,0.73,0.73}{##1}}}
\@namedef{PY@tok@c}{\let\PY@it=\textit\def\PY@tc##1{\textcolor[rgb]{0.24,0.48,0.48}{##1}}}
\@namedef{PY@tok@cp}{\def\PY@tc##1{\textcolor[rgb]{0.61,0.40,0.00}{##1}}}
\@namedef{PY@tok@k}{\let\PY@bf=\textbf\def\PY@tc##1{\textcolor[rgb]{0.00,0.50,0.00}{##1}}}
\@namedef{PY@tok@kp}{\def\PY@tc##1{\textcolor[rgb]{0.00,0.50,0.00}{##1}}}
\@namedef{PY@tok@kt}{\def\PY@tc##1{\textcolor[rgb]{0.69,0.00,0.25}{##1}}}
\@namedef{PY@tok@o}{\def\PY@tc##1{\textcolor[rgb]{0.40,0.40,0.40}{##1}}}
\@namedef{PY@tok@ow}{\let\PY@bf=\textbf\def\PY@tc##1{\textcolor[rgb]{0.67,0.13,1.00}{##1}}}
\@namedef{PY@tok@nb}{\def\PY@tc##1{\textcolor[rgb]{0.00,0.50,0.00}{##1}}}
\@namedef{PY@tok@nf}{\def\PY@tc##1{\textcolor[rgb]{0.00,0.00,1.00}{##1}}}
\@namedef{PY@tok@nc}{\let\PY@bf=\textbf\def\PY@tc##1{\textcolor[rgb]{0.00,0.00,1.00}{##1}}}
\@namedef{PY@tok@nn}{\let\PY@bf=\textbf\def\PY@tc##1{\textcolor[rgb]{0.00,0.00,1.00}{##1}}}
\@namedef{PY@tok@ne}{\let\PY@bf=\textbf\def\PY@tc##1{\textcolor[rgb]{0.80,0.25,0.22}{##1}}}
\@namedef{PY@tok@nv}{\def\PY@tc##1{\textcolor[rgb]{0.10,0.09,0.49}{##1}}}
\@namedef{PY@tok@no}{\def\PY@tc##1{\textcolor[rgb]{0.53,0.00,0.00}{##1}}}
\@namedef{PY@tok@nl}{\def\PY@tc##1{\textcolor[rgb]{0.46,0.46,0.00}{##1}}}
\@namedef{PY@tok@ni}{\let\PY@bf=\textbf\def\PY@tc##1{\textcolor[rgb]{0.44,0.44,0.44}{##1}}}
\@namedef{PY@tok@na}{\def\PY@tc##1{\textcolor[rgb]{0.41,0.47,0.13}{##1}}}
\@namedef{PY@tok@nt}{\let\PY@bf=\textbf\def\PY@tc##1{\textcolor[rgb]{0.00,0.50,0.00}{##1}}}
\@namedef{PY@tok@nd}{\def\PY@tc##1{\textcolor[rgb]{0.67,0.13,1.00}{##1}}}
\@namedef{PY@tok@s}{\def\PY@tc##1{\textcolor[rgb]{0.73,0.13,0.13}{##1}}}
\@namedef{PY@tok@sd}{\let\PY@it=\textit\def\PY@tc##1{\textcolor[rgb]{0.73,0.13,0.13}{##1}}}
\@namedef{PY@tok@si}{\let\PY@bf=\textbf\def\PY@tc##1{\textcolor[rgb]{0.64,0.35,0.47}{##1}}}
\@namedef{PY@tok@se}{\let\PY@bf=\textbf\def\PY@tc##1{\textcolor[rgb]{0.67,0.36,0.12}{##1}}}
\@namedef{PY@tok@sr}{\def\PY@tc##1{\textcolor[rgb]{0.64,0.35,0.47}{##1}}}
\@namedef{PY@tok@ss}{\def\PY@tc##1{\textcolor[rgb]{0.10,0.09,0.49}{##1}}}
\@namedef{PY@tok@sx}{\def\PY@tc##1{\textcolor[rgb]{0.00,0.50,0.00}{##1}}}
\@namedef{PY@tok@m}{\def\PY@tc##1{\textcolor[rgb]{0.40,0.40,0.40}{##1}}}
\@namedef{PY@tok@gh}{\let\PY@bf=\textbf\def\PY@tc##1{\textcolor[rgb]{0.00,0.00,0.50}{##1}}}
\@namedef{PY@tok@gu}{\let\PY@bf=\textbf\def\PY@tc##1{\textcolor[rgb]{0.50,0.00,0.50}{##1}}}
\@namedef{PY@tok@gd}{\def\PY@tc##1{\textcolor[rgb]{0.63,0.00,0.00}{##1}}}
\@namedef{PY@tok@gi}{\def\PY@tc##1{\textcolor[rgb]{0.00,0.52,0.00}{##1}}}
\@namedef{PY@tok@gr}{\def\PY@tc##1{\textcolor[rgb]{0.89,0.00,0.00}{##1}}}
\@namedef{PY@tok@ge}{\let\PY@it=\textit}
\@namedef{PY@tok@gs}{\let\PY@bf=\textbf}
\@namedef{PY@tok@ges}{\let\PY@bf=\textbf\let\PY@it=\textit}
\@namedef{PY@tok@gp}{\let\PY@bf=\textbf\def\PY@tc##1{\textcolor[rgb]{0.00,0.00,0.50}{##1}}}
\@namedef{PY@tok@go}{\def\PY@tc##1{\textcolor[rgb]{0.44,0.44,0.44}{##1}}}
\@namedef{PY@tok@gt}{\def\PY@tc##1{\textcolor[rgb]{0.00,0.27,0.87}{##1}}}
\@namedef{PY@tok@err}{\def\PY@bc##1{\setlength{\fboxsep}{-\fboxrule}\fcolorbox[rgb]{1.00,0.00,0.00}{1,1,1}{\strut ##1}}}
\@namedef{PY@tok@kc}{\let\PY@bf=\textbf\def\PY@tc##1{\textcolor[rgb]{0.00,0.50,0.00}{##1}}}
\@namedef{PY@tok@kd}{\let\PY@bf=\textbf\def\PY@tc##1{\textcolor[rgb]{0.00,0.50,0.00}{##1}}}
\@namedef{PY@tok@kn}{\let\PY@bf=\textbf\def\PY@tc##1{\textcolor[rgb]{0.00,0.50,0.00}{##1}}}
\@namedef{PY@tok@kr}{\let\PY@bf=\textbf\def\PY@tc##1{\textcolor[rgb]{0.00,0.50,0.00}{##1}}}
\@namedef{PY@tok@bp}{\def\PY@tc##1{\textcolor[rgb]{0.00,0.50,0.00}{##1}}}
\@namedef{PY@tok@fm}{\def\PY@tc##1{\textcolor[rgb]{0.00,0.00,1.00}{##1}}}
\@namedef{PY@tok@vc}{\def\PY@tc##1{\textcolor[rgb]{0.10,0.09,0.49}{##1}}}
\@namedef{PY@tok@vg}{\def\PY@tc##1{\textcolor[rgb]{0.10,0.09,0.49}{##1}}}
\@namedef{PY@tok@vi}{\def\PY@tc##1{\textcolor[rgb]{0.10,0.09,0.49}{##1}}}
\@namedef{PY@tok@vm}{\def\PY@tc##1{\textcolor[rgb]{0.10,0.09,0.49}{##1}}}
\@namedef{PY@tok@sa}{\def\PY@tc##1{\textcolor[rgb]{0.73,0.13,0.13}{##1}}}
\@namedef{PY@tok@sb}{\def\PY@tc##1{\textcolor[rgb]{0.73,0.13,0.13}{##1}}}
\@namedef{PY@tok@sc}{\def\PY@tc##1{\textcolor[rgb]{0.73,0.13,0.13}{##1}}}
\@namedef{PY@tok@dl}{\def\PY@tc##1{\textcolor[rgb]{0.73,0.13,0.13}{##1}}}
\@namedef{PY@tok@s2}{\def\PY@tc##1{\textcolor[rgb]{0.73,0.13,0.13}{##1}}}
\@namedef{PY@tok@sh}{\def\PY@tc##1{\textcolor[rgb]{0.73,0.13,0.13}{##1}}}
\@namedef{PY@tok@s1}{\def\PY@tc##1{\textcolor[rgb]{0.73,0.13,0.13}{##1}}}
\@namedef{PY@tok@mb}{\def\PY@tc##1{\textcolor[rgb]{0.40,0.40,0.40}{##1}}}
\@namedef{PY@tok@mf}{\def\PY@tc##1{\textcolor[rgb]{0.40,0.40,0.40}{##1}}}
\@namedef{PY@tok@mh}{\def\PY@tc##1{\textcolor[rgb]{0.40,0.40,0.40}{##1}}}
\@namedef{PY@tok@mi}{\def\PY@tc##1{\textcolor[rgb]{0.40,0.40,0.40}{##1}}}
\@namedef{PY@tok@il}{\def\PY@tc##1{\textcolor[rgb]{0.40,0.40,0.40}{##1}}}
\@namedef{PY@tok@mo}{\def\PY@tc##1{\textcolor[rgb]{0.40,0.40,0.40}{##1}}}
\@namedef{PY@tok@ch}{\let\PY@it=\textit\def\PY@tc##1{\textcolor[rgb]{0.24,0.48,0.48}{##1}}}
\@namedef{PY@tok@cm}{\let\PY@it=\textit\def\PY@tc##1{\textcolor[rgb]{0.24,0.48,0.48}{##1}}}
\@namedef{PY@tok@cpf}{\let\PY@it=\textit\def\PY@tc##1{\textcolor[rgb]{0.24,0.48,0.48}{##1}}}
\@namedef{PY@tok@c1}{\let\PY@it=\textit\def\PY@tc##1{\textcolor[rgb]{0.24,0.48,0.48}{##1}}}
\@namedef{PY@tok@cs}{\let\PY@it=\textit\def\PY@tc##1{\textcolor[rgb]{0.24,0.48,0.48}{##1}}}

\def\PYZus{\char`\_}

\def\PYZsh{\char`\#}

\def\PYZhy{\char`\-}
\def\PYZsq{\char`\'}
\def\PYZdq{\char`\"}

\makeatother

\newcommand\BibTeX{{\rmfamily B\kern-.05em \textsc{i\kern-.025em b}\kern-.08em
T\kern-.1667em\lower.7ex\hbox{E}\kern-.125emX}}

\title{A Community-Developed Domain Ontology for Magnetic Materials}

\providecommand{\keywords}[1]{\par\noindent\textbf{Keywords: }#1}

\author[1]{Wilfried Hortschitz\thanks{Corresponding author: Wilfried Hortschitz (wilfried.hortschitz@donau-uni.ac.at).}}
\author[1,2]{Santa Pile}
\author[1]{Claas Fillies}
\author[1]{Harald Oezelt}
\author[1]{Alexander Kovacs}
\author[3,4,5]{Hans Fangohr}
\author[3,4]{Samuel J. R. Holt}
\author[3,4]{Martin Lang}
\author[3,4]{Andrea Petrocchi}
\author[3,4]{Swapneel Pathak}
\author[3,4]{Michael P. Adams}
\author[6]{Jonas Winkler}
\author[6]{Thomas G. Woodcock}
\author[7]{William Rigaut}
\author[7]{Pierre Le Berre}
\author[7]{Nora M. Dempsey}
\author[8]{Alena Vishina}
\author[8]{M. Nur Hasan}
\author[8]{Georgia A. Marchant}
\author[8]{Heike C. Herper}
\author[1]{Thomas Schrefl}

\affil[1]{Department of Integrated Sensor Systems, University for Continuing Education Krems, Austria}
\affil[2]{Solid State Physics Department, Johannes Kepler University Linz, Austria}
\affil[3]{Max Planck Institute for the Structure and Dynamics of Matter, Hamburg, Germany}
\affil[4]{Center for Free-Electron Laser Science, Hamburg, Germany}
\affil[5]{University of Southampton, Southampton, UK}
\affil[6]{Leibniz Institute for Solid State and Materials Research Dresden, Germany}
\affil[7]{Univ. Grenoble Alpes, CNRS, Grenoble INP, Institut Néel, Grenoble, France}
\affil[8]{Department of Physics and Astronomy, Uppsala University, Uppsala, Sweden}

\date{} % arXiv typically omits a date

\begin{document}

\maketitle

\begin{abstract}
Magnetic materials play a crucial role in energy-related technologies, mobility, and sensing, but their complex multiscale behaviour and the coexistence of multiple unit systems pose persistent challenges for data exchange and interpretation. This paper presents a domain ontology for magnetic materials, developed within the European Union funded Magnetic Multiscale Modelling Suite (MaMMoS) project and aligned with the Elementary Multiperspective Material Ontology (EMMO). The ontology formalises intrinsic, hysteretic, and microstructural properties across multiple length scales and supports semantic interoperability between simulation tools, databases, and experimental workflows. One key feature of the ontology is its code-based and human-readable structure, enabled through the EMMOntoPy framework, which allows for direct manipulation and versioning without relying on opaque .owl or .ttl files. This facilitates collaborative development and improves transparency. The ontology supports FAIR (Findable, Accessible, Interoperable, Reusable) principles and is openly available for extension by the community. The aim of the MaMMoS project is to foster reproducibility, improve traceability, and promote the adoption of ontology in the magnetism domain. The ontology that has been developed already serves as the foundation for multiple software tools that handle all kinds of magnetic material data.
\end{abstract}

% Semantic interoperability, Materials informatics, 

\section{Introduction}

Magnetic materials are central to modern energy-, mobility-, and sensing-technologies; their complex multiscale behaviour and the coexistence of diverse unit systems continue to hinder consistent data exchange, interpretation, and reuse \citep{otte_bfo_2022, stier_materials_2024, horsch_applications_2021}. This paper introduces a domain ontology for magnetic materials or Magnetic Materials Ontology (MagMO), developed within the European Union funded  "Magnetic Multiscale Modelling Suite" project \citep{website_MaMMoS}

\cite{GitHub_repository_of_EMMO}, the Elementary Multiperspective Material Ontology, is a physics-grounded top-level ontology explicitly designed to provide a common, formally axiomatised framework for materials, processes, properties, and data across applied sciences, thereby enabling semantic interoperability and FAIR (Findable, Accessible, Interoperable, Reusable) data reuse. EMMO is described as a standard representational framework that captures fundamental concepts of physics, chemistry, and materials science, and is intended to be specialised by domain and application ontologies. \cite{keod24} show how the multiperspective architecture of EMMO (reductionistic, holistic, persistence, contrast, structural, and semiotic perspectives) can be reused systematically to build consistent domain ontologies (e.g., CHAMEO) and application ontologies (e.g., Battery Testing Ontology), demonstrating its versatility and interoperability benefits across heterogeneous material workflows. In the broader ontology landscape, \cite{lambrix_materials_2024} identify EMMO as a physics- and analytic-philosophy-based top-level ontology suitable as a foundation for domain ontologies such as the Materials Design Ontology, underlining its role as a reusable upper layer for materials informatics. A recent survey of ontologies in materials science and engineering by \cite{norouzi_sack_landscape_2024} further classifies EMMO, alongside other top-level ontologies, as a high-quality foundational ontology for representing key aspects of the field and guiding ontology selection. Moreover, performance evaluations in materials science by \cite{beygi_nasrabadi_performance_2025} have found that EMMO offers high semantic richness and domain coverage when used as the upper-level ontology for developing domain and application ontologies (e.g., for Brinell hardness testing), even though its detailed structure can affect query efficiency and integration in some contexts, highlighting both its expressive power and considerations for practical use.

Before developing a new ontology, an extensive review of existing resources was conducted, and no domain ontology specifically addressing magnetic materials could be identified. At the same time, the EMMO framework already provided foundational concepts such as coercivity and magnetic field strength, along with their associated SI units, which supported the reuse of established definitions and guided their integration into the implementation.
While the very recent study of \citep{Bekemeier2026_Magnetocaloric_Materials} has introduced an ontology for magnetocaloric materials, this work focuses on a distinct subset of magnetic phenomena and follows a different modelling approach, highlighting both the timeliness of ontology development in this field and the absence of a more general framework for magnetic materials.
\\
In MagMO, intrinsic, hysteretic, and microstructural properties were formalised across multiple length scales, enabling semantic clarity and interoperability between simulation, characterisation, and data management tools. It addresses long-standing issues in magnetism related to inconsistent units and ambiguous terminology, also offering a structured framework for unit conversion and data integration. Built using the Python library \cite{GitHub_repository_of_EMMOntoPy} and validated with EMMOCheck, both developed by the EMMC, the ontology supports FAIR principles \citep{wilkinson_fair_2016, jacobsen_fair_2024} and is openly available on multiple platforms for extension by the community, such as the code hosting platform GitHub \citep{GitHub_repository_MagneticMaterialsOntology}, and the research archive Zenodo \citep{fangohr_mammos-projectmagneticmaterialsontology_2024}, as well as a project-related own GitLab repository. This work aims to promote ontology adoption in magnetism, improve the traceability of models and data, and foster collaboration across disciplines.\\
Ontologies are important for making data reusable because they clarify the intended meaning and context of available data. As in many other fields of science and engineering, ontologies are necessary in materials research to preserve knowledge \citep{bayerlein_perspective_2022}.\\
For magnetism, ontologies offer a structured approach to managing the complexity of magnetic materials data by standardising definitions, units, and contextual metadata. They help prevent errors arising from inconsistent unit systems—such as SI versus CGS—by enforcing unit validation, enabling reliable conversions, and clarifying distinctions between related properties, such as volume and mass susceptibility. Consider the common practice of specifying the strength of the magnetic field \textit{H} in tesla. Although this usage is favoured for its graspable numerical values (typically 0.1–3~T), it introduces an ontological category error. Strictly speaking, tesla measures magnetic flux density (\textit{B}), while the ISO standard reserves the ampere per metre (A/m) for magnetic field strength (\textit{H}).
In data integration and sharing, ontologies ensure semantic consistency, reducing the risk of misinterpretation. This facilitates the accurate comparison, retrieval, and analysis of magnetic properties across diverse datasets and research environments.
In the field of magnetic materials, multiple unit systems and competing definitions have been used for decades. This lack of standardisation has long been recognised as a source of confusion. As Bennett \citep{bennett_comments_1978} already noted, ‘In order to apply SI units in the field of magnetism with a minimum of confusion, agreement and uniformity in symbols and definitions would be extremely helpful’.\\
Ontologies in the technical sciences are not limited to enforcing a physical unit system. Rather, they provide explicit definitions of variables and quantities so that data can be interpreted and, where appropriate, converted between unit systems in a well-defined way.

\section{Ontologies in Materials Science}
In various domains of materials science, ontologies have proven to be effective by enabling consistent representation and integration of complex data \citep{de_baas_friis_Goldbeck_review_2023, eisenbart_kupferdigital_2025, rajamohan_materials_2025, valdestilhas_intersection_2023, otte_bfo_2022}. In crystallography, they support standardised descriptions of lattice structures and symmetry operations. In computational materials design, ontologies facilitate interoperability between simulation tools and databases by aligning property definitions and units. Projects such as the Materials Data Science Ontology (MDS‑Onto, \citep{rajamohan_materials_2025}), the Materials Design Ontology (MDO \citep{lambrix_materials_2024}), and several EMMO‑powered initiatives by the European Materials Modelling Council \citep{EMMC} have demonstrated how ontologies improve data discoverability, reduce ambiguity, and support automated reasoning across heterogeneous datasets.\\
Ontologies in materials science provide a structured and machine-readable framework for describing complex materials, processes, and data, enabling seamless interoperability and integration across diverse datasets and tools. This standardisation enables and facilitates the integration and analysis of vast amounts of materials data, supporting, in combination with AI, more efficient and accurate predictions and discoveries. Additionally, ontologies enhance collaboration by allowing researchers to share and understand each other's work much more effectively. 

EMMO, developed by the European Materials Modelling Council, stands out for its rigorous, physics-based semantic framework that ensures an unambiguous representation of material properties and their interrelations. Unlike other ontologies, EMMO focuses on aligning materials modelling with fundamental physical principles, bridging the gap between experimental data and computational models. This alignment enhances data reproducibility, facilitates cross-disciplinary collaboration, and supports the FAIR principles in research more comprehensively. By providing a standardised framework for representing and sharing knowledge in modern materials science, EMMO ensures that data is categorised in a consistent manner, enhancing its findability. The ontology uses unique identifiers (URIs, IRIs) for concepts and relationships, making research data easier to discover and connect across different projects. Additionally, EMMO is freely available not only on the project's GitHub page \citep{GitHub_repository_of_EMMO} and on platforms like MatPortal \citep{MatPortal_EMMO} but also via \href{https://w3id.org/emmo}{https://w3id.org/emmo}, which provides greater visibility and access to the ontology within the research community.
The physics-based semantic framework provided by EMMO \citep{horsch_applications_2021} aligns well with the needs for magnetic materials modelling. It supports the representation of materials, processes, and properties in a consistent and extensible manner.\\
EMMO also supports accessibility by being an open and transparent framework that researchers can freely access, modify, and build upon. As an open-source resource, EMMO encourages open science and enables easy retrieval and use, in line with the accessibility goals of the FAIR principles. The ontology fosters interoperability through the use of formalised machine-readable structures such as OWL (Web Ontology Language).\\
While the standard EMMO is comprehensive, its significant axiomatic density can present challenges for practical implementation; consequently, modularised alternatives, such as the 'EMMO-LITE' version, are currently being developed to improve usability and reduce latency during development and testing.\\
Finally, EMMO itself also promotes interoperability by reusing established ontologies, such as \citep{qudt}, ensuring consistency across disciplines and enhancing data exchange. Through these features, EMMO contributes significantly to making scientific knowledge in materials science more accessible, shareable, and reusable, in line with the core principles of the FAIR framework.

Beyond the foundational EMMO framework—primarily developed via graphical user interfaces such as Protégé \citep{Musen2015Protege}—the EMMC and its community provide the EMMOntoPy Python library to enhance interoperability. This library facilitates seamless integration across diverse platforms by providing robust tools for the parsing and serialisation of various ontological formats, including .owl, .ttl, and .rdf.

\begin{figure}
    \centering
    \includegraphics[width=12cm]{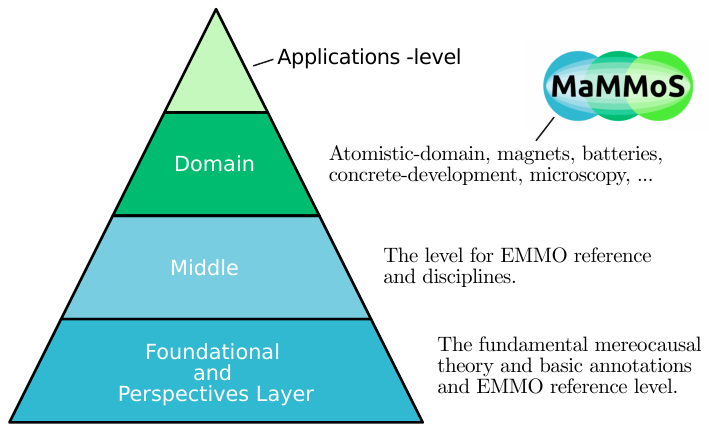}
    \caption{Hierarchy of ontologies in the EMMO ecosystem. The newly introduced domain-specific ontology for magnetic materials re-uses the existing Foundational (Top-) and Middle layer ontologies of EMMO. In the future ontologies for applications can also be built within the EMMO ecosystem.}
    \label{fig:hierarchy_of_EMMO_ecosystem}
\end{figure}

The EMMO ecosystem consists of a hierarchical structure as depicted in Figure~\ref{fig:hierarchy_of_EMMO_ecosystem}, with foundational and perspective-level ontologies, middle-level ontologies, domain-specific ontologies, and relevant ontologies for applications. The so-called "top-level" and middle levels of the EMMO define its foundational concepts and intermediate hierarchies, which provide the semantic structure to unify and connect domain-specific knowledge. Detailed insights on this topic can be found in the official EMMO GitHub repository \citep{GitHub_repository_of_EMMO}. 
This paper introduces the Magnetic Materials Ontology (MagMO), a domain ontology for magnetic materials developed within the European project Magnetic Multiscale Modelling Suite (MaMMoS) \citep{website_MaMMoS}. The ontology is designed to support unit standardisation, multiscale modelling, and semantic integration across magnetic materials data and software tools. The ontology created for magnetic materials will serve as a domain ontology, focusing on a specific area of research while adhering to the broader EMMO framework. All available ontologies within the EMMO ecosystem are designed to support one another, and their flexible structure ensures that they can be extended to create ontologies for specific applications across various fields, as seen in the ongoing development in the GitHub repository of EMMO.

Since the EMMO ecosystem is under continuous development, and in order to ensure version compatibility with the ontology for magnetic materials, one specific version of the first stable release of EMMO (1.0.2) was tested and stored locally in the GitHub repository of MaMMoS \citep{GitHub_repository_MagneticMaterialsOntology}.
Reusing EMMO makes it possible to concentrate on modelling and transferring the relevant domain knowledge—for example, clarifying what remanent magnetisation represents in the context of magnetic properties—without needing to redefine fundamental concepts such as magnetisation or basic units like the tesla.

\section{Magnetic Materials}
Magnetic materials are essential for many applications in energy, information, and communication technologies. Soft and hard magnetic materials, for example, are essential for power generation, energy conversion, and e-mobility \citep{cullity_introduction_2009, torta_exploring_2024, orlova_permanent_2024, coey_handbook_2021, coey_perspective_2020}. Their performance directly affects the efficiency of electric motors, generators, and transformers, which are key components in the transition to low-carbon technologies \citep{podmiljsak_future_2024, torta_exploring_2024}. For example, hysteresis losses in soft magnetic cores determine the thermal management needs and energy efficiency of electrical machines \citep{osemwinyen_core_2024}. In permanent magnet systems, material limitations constrain design flexibility and sustainability, especially due to reliance on rare-earth elements. Another application of these materials is the field of sensors, which are widely used in consumer and industrial applications such as current, wheel-speed, and position sensors in automotive applications. Permanent magnets also play a major role in optical image stabilisation systems for cameras in mobile phones \citep{slanovc_designing_2022, leitao_enhanced_2024}. The functional properties at the device level depend on the magnetic properties of the materials, for instance in electrical machines the properties of magnetic materials are core to their overall performance. In many motors, hysteresis losses in the soft magnetic stator core have a significant impact on the overall efficiency class, which in turn affects the effort required for thermal management. In permanent magnet synchronous motors, today’s hard ferrites and rare-earth magnets are limiting the motor designs and production processes. However, a novel soft magnetic material with textured magnetisation and hysteresis properties may also enable new motor designs with improved performance for specific applications (e.g.\ noise reduction).

Despite the importance of magnetic materials, the characterisation and modelling of magnetic materials remain fragmented. Properties such as magnetisation, coercivity, and susceptibility are often reported using inconsistent units and definitions, complicating data reuse and comparison \citep{bennett_comments_1978}. The coexistence of SI and CGS systems, along with varying interpretations of derived quantities, still leads to errors in both academic and industrial contexts, although this issue has been discussed for many years, as in \citep{goldfarb_units_1985}. This issue is particularly critical in multiscale modelling, where data must be integrated across atomistic, mesoscopic, and macroscopic levels.
Ontologies offer a structured solution to these challenges by providing formal, machine-readable representations of partly highly specialised domain knowledge. In materials science, ontologies have enabled semantic interoperability, improved data discoverability, and supported automated reasoning \citep{norouzi_sack_landscape_2024, himanen_data-driven_2019}. 
One key design goal of the MaMMoS project is to make the ontology directly accessible and editable by researchers, avoiding the complexity of traditional ontology formats and enabling version control through code-based workflows. This will be demonstrated in the section about the use case. %\ref{usecase}.

\begin{figure*}
    \centering
    \includegraphics[width=1\linewidth]{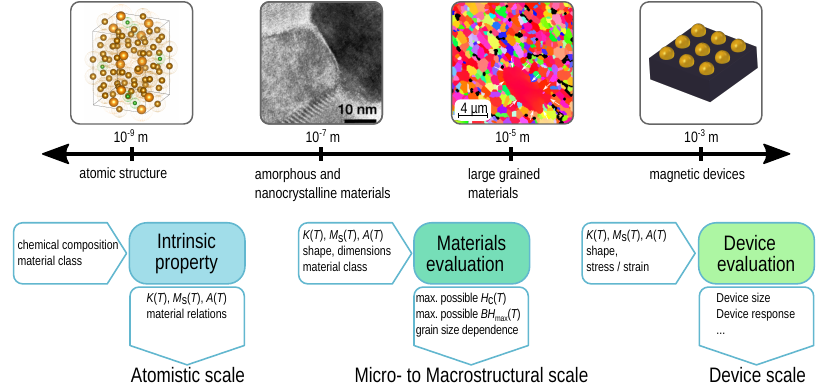}
    \caption{Bridging the length scales for magnetic materials modelling. MaMMoS provides a modelling and simulation suite for intrinsic property evaluation, materials evaluation, and device evaluation.}
    \label{fig:Bridging_length_scales}
\end{figure*}

The magnetic properties of materials arise from the interplay between quantum-physical electronic properties at the atomistic length scale and the material’s nanoscale and microscale structure (Figure~\ref{fig:Bridging_length_scales}). 
Due to their complexity, magnetic properties must be investigated across different length and time scales, which often limits the development of new magnetic materials and devices. The MaMMoS project aims to develop a magnetic multiscale modelling suite for the design and optimisation of magnetic materials and devices, combining multiscale modelling, characterisation, and numerical optimisation. The domain ontology for magnetic materials presented here was developed to achieve interoperability between simulation and analysis tools and to support the exchange of magnetic materials data across different length scales. Furthermore, it may serve as a key, openly available reference for terminology and units relevant to the description and exchange of magnetic materials properties in databases. The need for MagMO arises from the complexity of magnetic materials and their properties depicted here and briefly discussed in the following sections. MagMO aims to address:
\begin{itemize}
    \item The interplay between intrinsic magnetic properties of the material and the microstructure of the material, which governs the macroscopic hysteresis properties. 
    \item The multi-scale nature of magnetic materials, which is essential in the simulation and design of magnets with tailored magnetic properties.
    \item The physical quantities and the proper SI units used in magnetism.
\end{itemize}

For the development of the ontology for magnetic materials, MaMMoS uses data standards that comply with the standards and principles of Materials Modelling Data \citep{MODA} and Characterisation Data \citep{CHADA}, and with domain ontologies applied in the context of the EMMO.

By incorporating existing data from MODA and CHADA, the ontology ensures compatibility with existing materials-modelling frameworks. It is designed to support consistent terminology, facilitate data sharing, and enable interoperability between simulation and characterisation tools.

\subsection{Magnetic Properties}
\begin{figure}
    \centering
    \includegraphics[width=10cm]{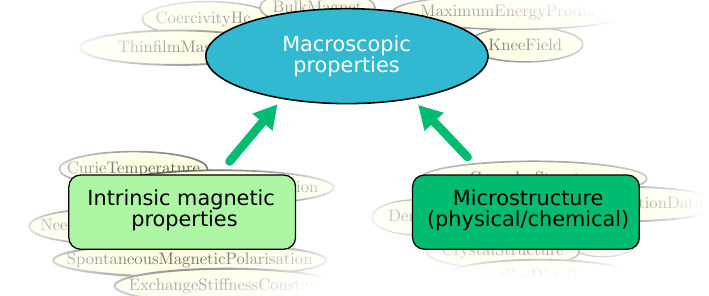}
    \caption{Basic categorisation of the properties of magnetic materials into three main groups.}
    \label{fig:Basic_categorisation_properties}
\end{figure}

From this section onwards, variables typeset in \texttt{monospace}, such as \texttt{MagneticMaterial}, correspond to the exact class names used in the ontology and therefore follow the PascalCase naming convention.
The intrinsic magnetic properties dependening on the temperature $T$, such as the \texttt{UniaxialAnisotropyConstant} $K_\mathrm{u}(T)$, the \texttt{SpontaneousMagnetization} $M_\mathrm{s}(T)$, and the \texttt{Exchange-\allowbreak StiffnessConstant} $A(T)$, arise from the crystal (atomic) structure of the material. In addition to the intrinsic magnetic properties, the nanoscale and microscale internal structure of the material plays an important role. Magnetic hysteresis or macroscopic properties such as the \texttt{RemanentMagneticPolarisation} $J_\mathrm{r}$, the \texttt{CoerciveField} $H_\mathrm{c}$, and \texttt{MaximumEnergyProduct} $(BH)_\mathrm{max}$ (see also Figure~\ref{fig:Parameters_necessary_for_the_description_of_a_hysteresis_loop}) depend both on the intrinsic magnetic properties as on the microstructure. Important microstructural properties that influence the macroscopic properties include the grain size, grain shape, and the properties and distribution of secondary phases. This dependence is schematically shown in Figure~\ref{fig:Basic_categorisation_properties}. Hysteresis properties are critical figures of merit for applications. For example, the coercive field quantifies the field scale for magnetisation reversal on a demagnetisation branch, while the maximum energy product is a technical design and comparison parameter for permanent magnet materials.

The related classes and attributes of the magnetic material ontology are described in the following section.

\subsection{Bridging the Length Scales}\label{sec:Bridg}
As mentioned earlier, magnetic properties of materials arise from the interplay between the quantum physics electronic properties on the atomistic length scale and the material’s nanoscale and microscale structure. The length scales addressed in the domain ontology are shown in Figure~\ref{fig:Bridging_length_scales}. The MaMMoS project develops data-driven models for magnetic materials at lengths ranging from the atomistic scale to device level through the fusion of characterisation and modelling data. Therefore, MaMMoS aims to link the existing software tools across the different length scales to establish standards for data exchange in applied magnetism and tailor the software suite according to European industry needs. In order to do that, it is necessary to make the interconnection between the magnetic properties on the different length scales clear.
By serving as a common semantic reference, the ontology resolves inherent ambiguities between disparate simulation frameworks, ensuring that parameters such as the 'exchange constant' are consistently mapped between density functional theory (DFT) and micromagnetic modelling. 
As an example of using the ontology for bridging length scales between DFT, spin dynamics, and micromagnetics, we refer to the MaMMoS software suite \citep{GitHub_repository_MaMMoS}, where the implemented tools explicitly use the ontology when passing parameters as described in detail by \citep{Fangohr_IEEE_2026_MaMMoS}.  

\subsection{Units for Magnetic Properties}

The overall aim is to connect data and software tools across different length scales, establish common standards for data exchange in applied magnetism, and ensure compatibility with industrial requirements. A central aspect of this work is the standardisation of units for magnetic properties. For this reason, the ontology for magnetic materials relies on SI units. % , as listed in Table 1

Different unit systems have historically been used in magnetism, and organisations such as the IEC, NIST, IEEE, and ISO provide extensive guidance on which units are recommended for specific contexts. Nevertheless, both software and hardware developers continue to supply key magnetic quantities in non‑SI units, such as the CGS unit Oersted for magnetic field strength. Fortunately, standards bodies including the IEEE, ISO, and the QUDT initiative offer well‑defined conversion factors to SI‑compliant units like ampere per metre (A/m). Since EMMO incorporates QUDT, it already contains these unit definitions and conversion rules, enabling semantically grounded and automated conversion to the correct SI units. This capability is also illustrated by the examples presented in the "Use Case Demonstration" section.

\section{MagMO}
This section describes the reasoning behind the chosen design of MagMO and explains how existing resources were reviewed, why EMMO was selected as the foundation, and how expert knowledge shaped the initial structure.

\subsection{Strategy for Ontology Development}
The development of the ontology began with a review of existing ontologies in the domain. Since no ontology specifically for magnetism was identified, the development process was initiated using EMMO, which was selected as the foundation due to the features discussed previously. The initial structure of the ontology was developed with input from domain experts. Information from workflows described with modelling data \citep{MaMMoS_MODA_initial_version} and characterisation data \citep{MaMMoS_CHADA_initial_version} documents for processes used in the MaMMoS-project was incorporated to ensure that all relevant properties were appropriately represented for all necessary length scales.

\begin{table}[ht]
\small
\centering
\caption{Strategy for building the MagMO ontology.}
\label{tab:strategy_for_building_the_ontology}
\begin{tabularx}{\textwidth}{p{0.43\textwidth} X}
\toprule
\textbf{Task} & \textbf{Method} \\
\midrule

Build an EMMO-based domain ontology &
Use EMMO 1.0.2 as top- and middle-level ontology \\

Include all relevant classes and attributes &
Cover the input and output of the simulation and measurement
workflows as depicted in the MODA diagrams (deliverable 4.1 \citep{MaMMoS_MODA_initial_version}) and the
CHADA diagrams (deliverable 3.1, \citep{MaMMoS_CHADA_initial_version}) \\

Respect current domain knowledge &
Use elucidations and comments in accordance with the definitions
given in widely accepted textbooks\textsuperscript{a} \\

Refer to community knowledge &
Make references to Wikipedia\textsuperscript{b},
Wikidata\textsuperscript{c}, and
Electropedia\textsuperscript{d}
wherever possible \\

Make it expandable and adaptable by members of the magnetism community &
Publish the Python code, which uses EMMOntoPy, publicly on GitHub \\

Make the ontology conform with EMMO conventions &
Check EMMO conventions with emmocheck (the check runs automatically
whenever a new version is pushed on GitHub) \\

Check the consistency of the ontology &
Run the HermiT reasoner to ensure the ontology's structure and data
are logically coherent \\

\bottomrule
\end{tabularx}

\vspace{4pt}
\begin{flushleft}
\footnotesize
\textsuperscript{a}The following textbooks were used: \cite{coey_magnetism_2009}, \cite{lacheisserie_magnetism_2002}, \cite{coey_Skomski_permanent_1999}, \cite{cullity_introduction_2009}.\\
\textsuperscript{b}https://en.wikipedia.org/ \quad
\textsuperscript{c}https://www.wikidata.org/ \quad
\textsuperscript{d}https://www.electropedia.org/
\end{flushleft}
\end{table}

\section{Structure of MagMO}\label{sec:Structure}
The structure reflects the scientific understanding that macroscopic magnetic behaviour arises from both intrinsic properties and the microstructure of the material. The ontology is therefore arranged into interconnected parts that represent intrinsic properties, magnetic hysteresis properties, and the physical or chemical microstructure.

\subsection{Magnetic Properties}
\begin{figure*}
    \centering
    \includegraphics[width=1\linewidth]{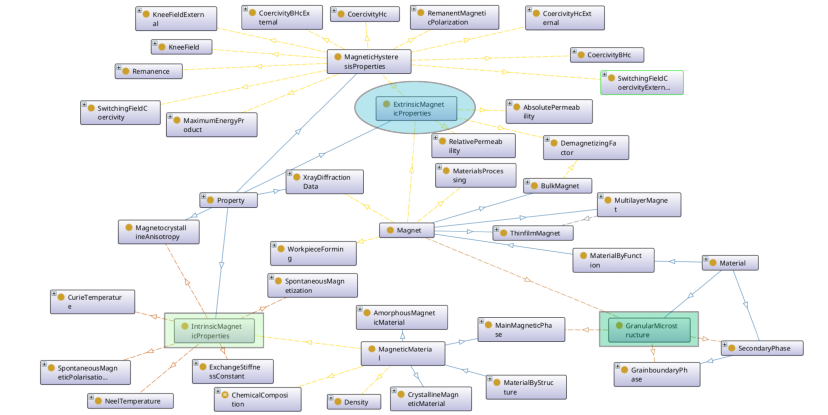}
    \caption{Three main categories of properties for magnetic materials together with the related elements in the ontology. This representation, exported from Protégé, serves to illustrate the clustering of the primary property groups. The figure is intended as a schematic overview of the mereological structure; consequently, certain labels are truncated to maintain visual clarity despite the high complexity of the underlying semantic interconnections.}
    \label{fig:Three_main_categories_of_prop_with_related_elements}
\end{figure*}

One entry point to the magnetic materials ontology reflects the interplay between intrinsic magnetic characteristics and the microstructure, which together give rise to the macroscopic properties of a magnet. This improves clarity and understanding when the use case focuses on the magnetic properties. Therefore, the ontology was organised into three main categories of properties:
\begin{enumerate}
    \item Intrinsic Magnetic Properties – e.g., spontaneous magnetisation, anisotropy, and exchange constant.
    \item Magnetic Hysteresis Properties or Macroscopic Properties – e.g., coercive field, remanence, and energy product.
    \item Physical/Chemical (Micro-) Structure – related to grain structure and material phases.
\end{enumerate}
This categorisation is shown in Figure~\ref{fig:Three_main_categories_of_prop_with_related_elements}, which reflects the understanding that macroscopic properties arise from intrinsic magnetic properties and the microstructure of the magnet (compare Figure~\ref{fig:Basic_categorisation_properties}).

\begin{figure*}
    \centering
    \includegraphics[width=1\linewidth]{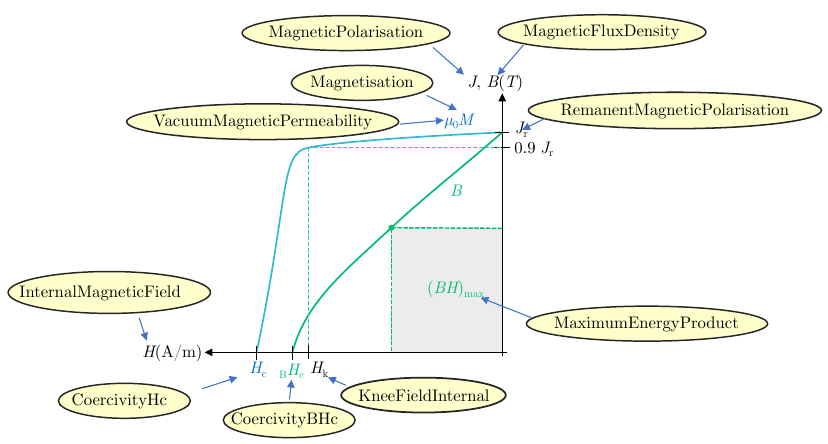}
    \caption{Parameters necessary for the description of a hysteresis loop and how they are named in the ontology.}
    \label{fig:Parameters_necessary_for_the_description_of_a_hysteresis_loop}
\end{figure*}

The essential practical characteristic of any magnetic material is the irreversible nonlinear response of magnetisation $M$ to an imposed magnetic field $H$. Based on the hysteresis loop, various macroscopic properties of the magnet can be defined. Figure~\ref{fig:Parameters_necessary_for_the_description_of_a_hysteresis_loop} shows the second quadrant of the hysteresis loop and how the main magnetic hysteresis (macroscopic) properties are obtained. The corresponding elements from the ontology are given in the oval shapes.

\begin{figure}
    \centering
    \includegraphics[width=1\linewidth]{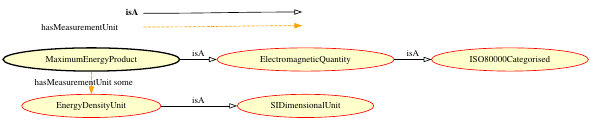}
    \caption{Quantities and their units are defined within the ontology. Here the relation to the unit of the maximum energy product via 'hasMeasurementUnit' is shown.}
    \label{fig:Quantities}
\end{figure}

The \texttt{MagneticPolarisation} $J$ is plotted as a function of the \texttt{InternalMagneticField} $H$. The magnetic polarisation is the product of the magnetisation times the \texttt{VacuumMagneticPermeability} $\mu_0$. The same graph also shows the \texttt{MagneticFluxDensity} $B(H)$. $B$ and $J$ are measured in tesla (T). The internal field is given in ampere per metre (A/m). The remanent magnetic polarisation $J_\mathrm{r}$ remains when the applied field is restored to zero during the hysteresis loop. The coercivity is the internal magnetic field -$H_\mathrm{c}$ at which the macroscopic magnetisation vanishes. The internal flux coercivity $_\mathrm{B}H_\mathrm{c}$ (\texttt{CoercivityBHc}) is defined as the internal field on the $B(H)$ loop at which $B = 0$. The maximum working field - also named \texttt{KneeFieldInternal} $H_\mathrm{k}$, is defined as the reverse internal field for which the magnetic polarisation is reduced by $10\%$; thus it corresponds to the point on the demagnetisation curve at which $J = 0.9~J_\mathrm{r}$. The maximum energy product $(BH)_{\mathrm{max}}$ equals the area of the largest second-quadrant rectangle which fits under the $B-H$ loop. The maximum energy product is the key figure of merit of a permanent magnet. It is twice the energy stored in the stray field of a magnet of optimal shape.
An important aspect of the ontology is the proper definition of quantities and their units. Figure \ref{fig:Quantities} shows the local ontology context of the maximum energy product, including adjacent classes and relations. Here, the magnetic materials ontology makes extensive use of the measurement units defined in the top- and middle-level ontology EMMO.

\subsubsection{A Magnet and its Constituent Parts:}
\begin{figure*}
    \centering
    \includegraphics[width=1\linewidth]{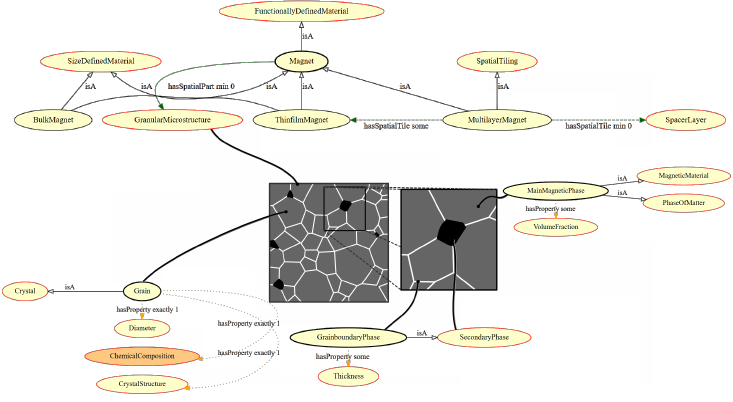}
    \caption{The definition of a magnet and its microstructure. The centre shows a sketch of the granular microstructure based on a scanning electron microscopy image of a permanent magnet.}
    \label{fig:Magnet_and_its_micorstructure}
\end{figure*}

A different perspective on the magnetic material ontology starts from the hierarchical structure of a magnet. It reveals its physical parts at different length scales. A magnet is a functionally defined material. Possible subclasses of a magnet are bulk magnet, thin film magnet, or multilayer magnet. A magnet may have a granular microstructure. The spatial parts of the granular microstructure are the main magnetic phase, the grain boundary phase, and secondary phases.
In Figure~\ref{fig:Magnet_and_its_micorstructure} the main magnetic phase is a subclass of magnetic material. It has the property "volume fraction". The granular microstructure consists of grains. A grain is a crystal. In an equiaxial case it has the following properties: a diameter, a chemical composition, and a crystal structure. It also has (not shown in Figure~\ref{fig:Magnet_and_its_micorstructure}) a crystallographic orientation or a grain misalignment angle.

\begin{figure*}
    \centering
    \includegraphics[width=1\linewidth]{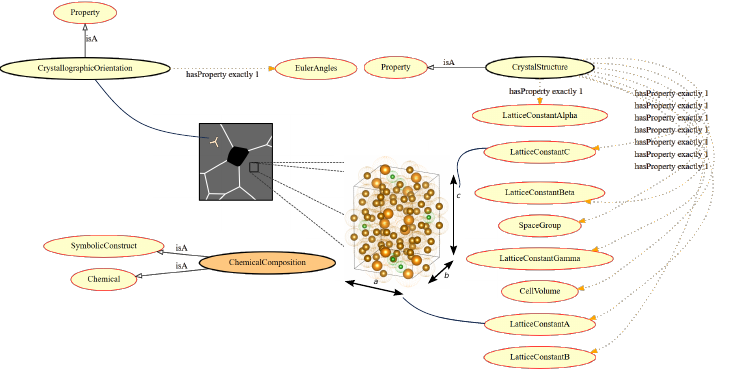}
    \caption{A grain has a crystallographic orientation, a chemical composition, and a crystal structure.}
    \label{fig:Grain_crystOrientation_chemComp}
\end{figure*}

The principle constituent parts of the granular structure are the grains. Each grain possesses a crystal structure, a chemical composition, and a crystallographic orientation. Figure~\ref{fig:Grain_crystOrientation_chemComp} illustrates the properties of a grain and depicts the entities of the ontology at the atomistic length scale.

\begin{figure*}
    \centering
    \includegraphics[width=1\linewidth]{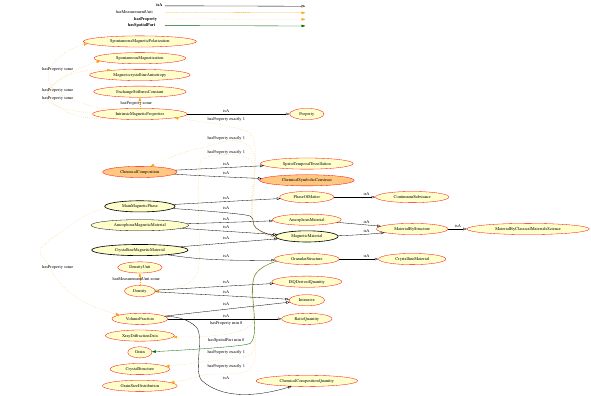}
    \caption{Visualisation of the hierarchical structure surrounding the entry MagneticMaterial in the ontology.}
    \label{fig:Vis_hierarchical_structure}
\end{figure*}

Figure~\ref{fig:Vis_hierarchical_structure} depicts the hierarchical structure with up to two neighbouring elements surrounding the main entry magnetic material in the ontology. A magnetic material has intrinsic magnetic properties and a chemical composition. A magnetic material can be amorphous or crystalline. Therefore, two subclasses, amorphous magnetic material and crystalline magnetic materials, are defined. A crystalline magnetic material is a granular structure. Properties of the granular structure are a crystal structure and a grain size distribution. The entry for X-ray diffraction is intended to describe measured data.

At the time of writing, the ontology for magnetic materials comprised 118 uniquely defined classes, along with nine classes reused from the EMMO domain‑crystallography ontology\footnote{https://github.com/emmo-repo/domain-crystallographys}. These reused elements, such as CellVolume, CrystalStructure, and LatticeParameter, were incorporated directly because the domain‑crystallography ontology is still based on an earlier EMMO 1.0.0‑beta release. Integrating it without modification would introduce version inconsistencies with the current stable EMMO version adopted in this work.
In addition to the core classes, the ontology includes more fine‑grained variants of selected magnetic quantities, reflecting their different uses in modelling and characterisation workflows. For example, five forms of coercivity (e.g., intrinsic, extrinsic, flux‑related-internal, flux‑related-external, local) and three forms of loop squareness are provided. These variants allow users to select the definitions that match their specific modelling requirements or measurement conventions, supporting precise and context‑appropriate descriptions of magnetic behaviour.

% Using build_onto.py (branch Energy-terms), I counted ontology terms as class declarations inside the with onto: block.

% Total entries declared in this ontology: 127
% Reused/copied from domain-crystallography (practical match, including renamed aliases): 9
% Unique to your magnetic materials ontology (127 - 9): 118
% The 9 reused crystallography concepts are:

% SpaceGroup (case variant of Spacegroup)
% CellVolume
% CrystalStructure
% LatticeParameterA/B/C/Alpha/Beta/Gamma concepts (in your file as LatticeConstant*, with altLabel set to LatticeParameter*)

% Coercivity variants: 5 strict Coercivity* classes
% CoercivityHc
% CoercivityBHc
% CoercivityHcExternal
% CoercivityBHcExternal
% LocalCoercivity

% For LoopSquareness variants: 3
% LoopSquarenessFactorInternal
% LoopSquarenessFactorExternal
% LoopSquareness

\subsection{Implementation and Tools}
The EMMOntoPy library developed by EMMC provides a powerful Python-based framework for creating and managing ontologies, with particular support for EMMO-based workflows. One of the key advantages of using EMMOntoPy for ontology creation over manually editing files such as .ttl (Turtle) is the benefit of version control, e.g.\ via GitHub. With traditional approaches, where ontologies are directly written in Turtle or RDF formats, the use of unique identifiers (URIs, IRIs) makes the files hard to interpret, especially when dealing with large ontologies. As these identifiers grow, the readability of the files diminishes, making it difficult for users to understand, track, and modify the ontology effectively. EMMOntoPy solves this by allowing users to generate and manipulate the ontology programmatically through Python scripts. This approach not only enhances version control by enabling systematic tracking of changes, but also supports the use of modular, reusable components and consistent updates, making the ontology easier to maintain over time. Furthermore, using Python code for ontology creation facilitates easier collaboration, especially when sharing on platforms like GitHub, as others can easily contribute by adding new classes, relations, or modifications through version-controlled pull requests, ensuring a smooth and efficient workflow for collaborative ontology development.
Moreover, EMMOntoPy includes EMMOCheck and OntoDoc as essential tools for ontology validation and documentation. EMMOCheck is an automated tool that checks the integrity of the ontology, ensuring it adheres to specified constraints and guidelines. This tool can help identify inconsistencies and errors like missing labels or definitions of units before deploying or sharing the ontology, which is crucial for maintaining quality in complex scientific ontologies. Furthermore, the integration of the EMMOCheck utility within the development pipeline ensures strict adherence to EMMO-compliance criteria by automatically verifying that all classes possess mandatory elucidations and that physical quantities are correctly aligned with the QUDT framework. EMMODoc generates comprehensive documentation from the ontology, making it easier for users to understand the structure, relationships, and definitions within the ontology. This feature greatly enhances the usability of the ontology, ensuring that users can quickly get up to speed and utilise the ontology in their work. Together, these tools help streamline the development, maintenance, and use of the ontology, offering clear advantages over traditional, manually edited ontology files. To generate images related to entries in the ontology for magnetic materials, a modified version of OntoDoc, called MaMMoSDoc, was developed. A step-by-step guide on how to use MaMMoSDoc can be found alongside the shared code in the ‘doc’ subfolder of the MaMMoS GitHub repository. The documentation of MagMO\footnote{https://emmo-repo.github.io/domain-magnetic-materials/}
was created with the above-mentioned tools.

Finally, Protégé \citep{Musen2015Protege}, a widely used and actively maintained graphical tool, was used during the development. It is favoured by the majority of the ontology community due to its free and open-source nature. During the development, Protégé was initially utilised not only for visualising the ontology but also for running the reasoner and exporting an inferred version, enhancing the overall development and validation process. Furthermore, the integrated plug-ins such as OntoGraf were used to generate visual representations of subsets of the ontology for this paper.
It was decided to use the latest stable version 1.0.2 of EMMO because of its extended content and bug fixes compared to previous versions.
The MaMMoS GitHub repository provides a step-by-step description of how to view and use the ontology. Furthermore, information is also provided on how to update and run the underlying Python code to recreate the final .ttl ontology file.

\section{Use Case Demonstration}\label{usecase}

Before presenting practical examples of how the ontology can be used, we briefly illustrate how entries are defined in the Python‑based implementation using the EMMOntoPy library. The following code excerpt shows the definition of \texttt{SpontaneousMagnetization} in the main ontology‑construction file \texttt{build\_onto.py}\footnote{https://github.com/MaMMoS-project/MagneticMaterialsOntology/blob/main/src/build\_onto.py}. The class inherits from the EMMO element \texttt{emmo.ElectromagneticQuantity}, and its labels, alternative names, relations, and external references are expressed in a clear and human‑readable form.
To support term lookup across English variants, MagMO uses US English as the preferred label and provides British English spellings as alternative labels (e.g., polarization vs. polarisation).
\begingroup
\sloppy

% Reduced font size + safe code-breaking behaviour
\fvset{breaklines=true, breakanywhere=true, fontsize=\scriptsize}

\tcbset{
  usecasecell/.style={
    breakable,
    size=fbox,
    width=\linewidth,
    boxrule=1pt,
    pad at break*=1mm,
    colback=cellbackground,
    colframe=cellborder,
    left=0.5mm, right=0.5mm, top=0.5mm, bottom=0.5mm,
    boxsep=1mm
  }
}

\begin{minipage}{\linewidth}
\begin{tcolorbox}[usecasecell]
\prompt{In}{incolor}{ }{\boxspacing}
\begin{Verbatim}[commandchars=\\\{\}]
\PY{k+kd}{class} \PY{n+nc}{SpontaneousMagnetization}\PY{p}{(}\PY{n}{emmo}\PY{o}{.}\PY{n}{ElectromagneticQuantity}\PY{p}{)}\PY{p}{:}
    \PY{l+s+sd}{"""The spontaneous magnetization, Ms, of a ferromagnet is the result}
\PY{l+s+sd}{of alignment of the magnetic moments of individual atoms. Ms exists}
\PY{l+s+sd}{within a domain of a ferromagnet."""}

    \PY{n}{prefLabel} \PY{o}{=} \PY{n}{enUS}\PY{p}{(}\PY{l+s+s2}{"SpontaneousMagnetization"}\PY{p}{)}
    \PY{n}{altLabel} \PY{o}{=} \PY{p}{[}
        \PY{n}{enGB}\PY{p}{(}\PY{l+s+s2}{"SpontaneousMagnetisation"}\PY{p}{)}, 
        \PY{n}{pl}\PY{p}{(}\PY{l+s+s2}{"Ms"}\PY{p}{)}, 
    \PY{p}{]}

    \PY{n}{is\PYZus{}a} \PY{o}{=} \PY{p}{[}
        \PY{n}{emmo}\PY{o}{.}\PY{n}{hasMeasurementUnit}\PY{o}{.}\PY{n}{some}\PY{p}{(}\PY{n}{emmo}\PY{o}{.}\PY{n}{MagneticFieldStrengthUnit}\PY{p}{)}
    \PY{p}{]}

    \PY{n}{IECEntry} \PY{o}{=} \PY{n}{pl}\PY{p}{(}
        \PY{l+s+s2}{"https://www.electropedia.org/iev/iev.nsf/display?openform\\&amp;ievref=221-02-41"}
    \PY{p}{)}

    \PY{n}{wikipediaReference} \PY{o}{=} \PY{n}{pl}\PY{p}{(}
        \PY{l+s+s2}{"https://en.wikipedia.org/wiki/Spontaneous_magnetization"}
    \PY{p}{)}
\end{Verbatim}
\end{tcolorbox}
\end{minipage}

\endgroup

To demonstrate the practical use of the ontology within the framework, and in the spirit of applied ontology as promoted by this journal, a set of example Jupyter notebooks was developed \citep{MaMMoS_examples_notebooks} to illustrate how the ontology can be queried and integrated into real analysis workflows.

These notebooks show how ontology entities can be queried, how metadata can be accessed, and how the ontology can be integrated into typical analysis workflows. The simplest example is included in the folder ‘example\_use\_case’ MagneticMaterials\footnote{https://github.com/MaMMoS-project/MagneticMaterialsOntology} in the public GitHub repository of the MagneticMaterialsOntology. The required Python libraries can be installed using standard package managers such as \texttt{pip}\footnote{https://pypi.org/}.

A central use case concerns the correct and consistent handling of units. Data may be provided using a variety of unit systems—for instance, the CGS unit Oersted—but internal processing should always be performed in SI units. The correct conversion factor is obtained directly from the ontology. The following first code example illustrates how an entity representing spontaneous magnetisation can be created, how its labels and unique identifiers (IRI) can be accessed, and how its associated elucidation is retrieved from the ontology. It is shown how this ontology entry is used programmatically through the mammos‑entity package, which provides convenient interfaces for accessing ontology labels, units, and metadata.

% Demo: using mammos_entity to link data to ontology terms
% Content extracted from demo_usage_ontology_via_mammos_entity.tex (Jupyter notebook export)
% Included from main.tex inside the Use case section.

% \begingroup
\sloppy

% Make notebook cells respect the current text area (SAGE class can be narrow)
\fvset{breaklines=true,breakanywhere=true,fontsize=\footnotesize}
\tcbset{usecasecell/.style={breakable, size=fbox, width=\linewidth, boxrule=1pt, pad at break*=1mm,
  colback=cellbackground, colframe=cellborder, left=0.5mm, right=0.5mm, top=1.5mm, bottom=0.5mm, boxsep=1.5mm}}
\tcbset{usecaseout/.style={breakable, size=fbox, width=\linewidth, boxrule=.5pt, pad at break*=1mm,
  opacityfill=0, colframe=cellborder, left=0.5mm, right=0.5mm, top=0.5mm, bottom=1.5mm, boxsep=1mm}}

%%%%%%%%%%%%%%%%%%%%%%%%%%%%%%%%%%%%%%%%%%%%%%%%%%%%%%%%%%%%%%%
% 1. Import block (no output, so single block)
%%%%%%%%%%%%%%%%%%%%%%%%%%%%%%%%%%%%%%%%%%%%%%%%%%%%%%%%%%%%%%%

\begin{minipage}{\linewidth}
\begin{tcolorbox}[usecasecell]
\prompt{In}{incolor}{ }{\boxspacing}
\begin{Verbatim}[commandchars=\\\{\}]
\PY{c+c1}{\PYZsh{} Importing the mammos\PYZhy{}entity module, which provides the Entity class and specific classes for magnetic properties}
\PY{k+kn}{import}\PY{+w}{ }\PY{n+nn}{mammos\PYZus{}entity}\PY{+w}{ }\PY{k}{as}\PY{+w}{ }\PY{n+nn}{me}
\end{Verbatim}
\end{tcolorbox}
\end{minipage}

%%%%%%%%%%%%%%%%%%%%%%%%%%%%%%%%%%%%%%%%%%%%%%%%%%%%%%%%%%%%%%%
% 2. Entity definition + output glued together
%%%%%%%%%%%%%%%%%%%%%%%%%%%%%%%%%%%%%%%%%%%%%%%%%%%%%%%%%%%%%%%
An ontology entity can be instantiated directly by providing a value, a unit, and an optional description.

\begin{minipage}{\linewidth}
\begin{tcolorbox}[usecasecell]
\prompt{In}{incolor}{ }{\boxspacing}
\begin{Verbatim}[commandchars=\\\{\}]
\PY{c+c1}{\PYZsh{} Definition using mammos\PYZhy{}entity which}
\PY{n}{me}\PY{o}{.}\PY{n}{Entity}\PY{p}{(}\PY{l+s+s2}{\PYZdq{}}\PY{l+s+s2}{SpontaneousMagnetization}\PY{l+s+s2}{\PYZdq{}}\PY{p}{,} \PY{l+m+mf}{0.7}\PY{p}{,}
               \PY{l+s+s2}{\PYZdq{}}\PY{l+s+s2}{MA/m}\PY{l+s+s2}{\PYZdq{}}\PY{p}{,}
               \PY{l+s+s2}{\PYZdq{}}\PY{l+s+s2}{Measured with technique ABC}\PY{l+s+s2}{\PYZdq{}}\PY{p}{)}
\end{Verbatim}
\end{tcolorbox}
\begin{tcolorbox}[usecaseout]
\prompt{Out}{outcolor}{ }{\boxspacing}
\begin{Verbatim}[commandchars=\\\{\}]
Entity(ontology\_label='SpontaneousMagnetization', value=np.float64(0.7),
unit='MA / m', description='Measured with technique ABC')
\end{Verbatim}
\end{tcolorbox}
\end{minipage}

%%%%%%%%%%%%%%%%%%%%%%%%%%%%%%%%%%%%%%%%%%%%%%%%%%%%%%%%%%%%%%%
% 3. Ms definition (no output)
%%%%%%%%%%%%%%%%%%%%%%%%%%%%%%%%%%%%%%%%%%%%%%%%%%%%%%%%%%%%%%%
For frequently used magnetic quantities, the package provides specialised classes such as Ms, which streamline the creation of domain entries.

\begin{minipage}{\linewidth}
\begin{tcolorbox}[usecasecell]
\prompt{In}{incolor}{ }{\boxspacing}
\begin{Verbatim}[commandchars=\\\{\}]
\PY{c+c1}{\PYZsh{} Shorter and more practical alternative, defining a variable Ms which holds all the information}
\PY{n}{Ms} \PY{o}{=} \PY{n}{me}\PY{o}{.}\PY{n}{Ms}\PY{p}{(}\PY{p}{[}\PY{l+m+mi}{150}\PY{p}{,} \PY{l+m+mi}{200}\PY{p}{,} \PY{l+m+mi}{310}\PY{p}{]}\PY{p}{,} \PY{l+s+s2}{\PYZdq{}}\PY{l+s+s2}{kA/m}\PY{l+s+s2}{\PYZdq{}}\PY{p}{,} \PY{n}{description}\PY{o}{=}\PY{l+s+s2}{\PYZdq{}}\PY{l+s+s2}{Measured with technique ABC}\PY{l+s+s2}{\PYZdq{}}\PY{p}{)}
\end{Verbatim}
\end{tcolorbox}
\end{minipage}

%%%%%%%%%%%%%%%%%%%%%%%%%%%%%%%%%%%%%%%%%%%%%%%%%%%%%%%%%%%%%%%
% 4. Access ontology_label + output glued
%%%%%%%%%%%%%%%%%%%%%%%%%%%%%%%%%%%%%%%%%%%%%%%%%%%%%%%%%%%%%%%

\begin{minipage}{\linewidth}
\begin{tcolorbox}[usecasecell]
\prompt{In}{incolor}{ }{\boxspacing}
\begin{Verbatim}[commandchars=\\\{\}]
\PY{c+c1}{\PYZsh{} Accessing the label of Ms from the ontology, which was before automatically linked to the variable name Ms}
\PY{n}{Ms}\PY{o}{.}\PY{n}{ontology\PYZus{}label}
\end{Verbatim}
\end{tcolorbox}
\begin{tcolorbox}[usecaseout]
\prompt{Out}{outcolor}{ }{\boxspacing}
\begin{Verbatim}[commandchars=\\\{\}]
\PYZsq{}SpontaneousMagnetization\PYZsq{}
\end{Verbatim}
\end{tcolorbox}
\end{minipage}

%%%%%%%%%%%%%%%%%%%%%%%%%%%%%%%%%%%%%%%%%%%%%%%%%%%%%%%%%%%%%%%
% 5. Access ontology_label_with_iri + output glued
%%%%%%%%%%%%%%%%%%%%%%%%%%%%%%%%%%%%%%%%%%%%%%%%%%%%%%%%%%%%%%%
Once created, the variable provides convenient access to the ontology label associated with the corresponding class. If needed, the full ontology identifier including its unique IRI can be retrieved as well.
\begin{minipage}{\linewidth}
\begin{tcolorbox}[usecasecell, breakable]
\prompt{In}{incolor}{ }{\boxspacing}
\begin{Verbatim}[commandchars=\\\{\}]
\PY{c+c1}{\PYZsh{} Accessing the label of Ms with unique IRI from the ontology}
\PY{n}{Ms}\PY{o}{.}\PY{n}{ontology\PYZus{}label\PYZus{}with\PYZus{}iri}
\end{Verbatim}
\end{tcolorbox}
\begin{tcolorbox}[usecaseout, breakable]
\prompt{Out}{outcolor}{ }{\boxspacing}
\begin{Verbatim}[commandchars=\\\{\}]
\PYZsq{}SpontaneousMagnetization https://w3id.org/emmo/domain/magnetic-
materials\PYZsh{}EMMO\PYZus{}032731f8-874d-5efb-9c9d-6dafaa17ef25\PYZsq{}
\end{Verbatim}
\end{tcolorbox}
\end{minipage}

%%%%%%%%%%%%%%%%%%%%%%%%%%%%%%%%%%%%%%%%%%%%%%%%%%%%%%%%%%%%%%%
% 6. Elucidation block (input + raw output glued)
%%%%%%%%%%%%%%%%%%%%%%%%%%%%%%%%%%%%%%%%%%%%%%%%%%%%%%%%%%%%%%%

The ontology also stores description, so-called elucidations, which can be queried to retrieve a short textual explanation of the physical quantity.

\begin{minipage}{\linewidth}
\begin{tcolorbox}[usecasecell]
\prompt{In}{incolor}{ }{\boxspacing}
\begin{Verbatim}[commandchars=\\\{\}]
\PY{c+c1}{\PYZsh{} Accessing the description/definition/elucidation of Ms directly from the ontology}
\PY{n+nb}{print}\PY{p}{(}\PY{n}{Ms}\PY{o}{.}\PY{n}{ontology}\PY{o}{.}\PY{n}{elucidation}\PY{p}{[}\PY{l+m+mi}{0}\PY{p}{]}\PY{p}{)}
\end{Verbatim}
\end{tcolorbox}
\begin{Verbatim}[commandchars=\\\{\}]
The spontaneous magnetization, Ms, of a ferromagnet is the result
of alignment of the magnetic moments of individual atoms. Ms exists
within a domain of a ferromagnet.
\end{Verbatim}

\end{minipage}

% \endgroup

In a second step, the data created using mammos‑entity can be exported and uploaded to external FAIR repositories such as NOMAD, enabling sharing, validation, and long‑term preservation.
The second example shows how software developed in the MaMMoS project can be used to upload and share data created using the mammos-entity package (as in the example above).

\begin{figure}[H]
    \centering
    \includegraphics[width=1\linewidth]{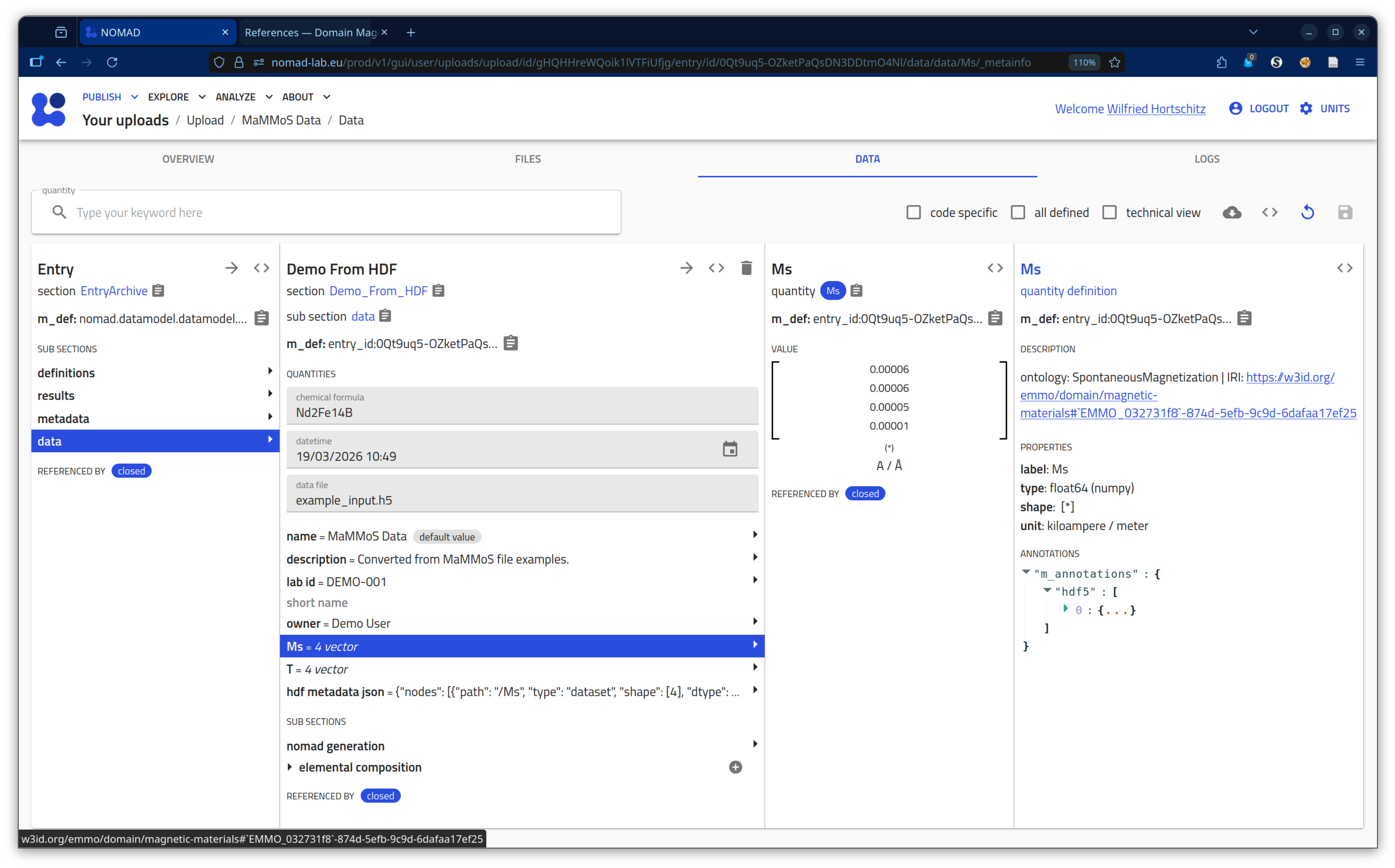}
    \caption{Data created using the mammos-entity Python library and uploaded to NOMAD, a platform for FAIR materials data management.}
    \label{fig:Nomad}
\end{figure}

Figure~\ref{fig:Nomad} shows data similar to the example above uploaded to the NOMAD\footnote{https://nomad-lab.eu} materials data management platform. In addition to the raw data and metadata, the platform shows an explanation of the quantity, here \texttt{SpontaneousMagnetization} for \texttt{Ms}, as well as a direct link to the publicly available online documentation of the corresponding ontology entry.\\
Together, these examples demonstrate how the ontology supports consistent variable definitions, correct SI conversions, and seamless integration into data‑processing and sharing workflows.

\section{Summary and Outlook}
The authors regard the presented ontology as an initial version. Experience from domain experts and feedback from colleagues working with the ontology will help to continuously adapt and improve it. We presented a few examples of how to easily use the ontology integrated into Python-based workflows to uniquely store data and share it according to FAIR principles.
Among other applications, we plan to use MagMO within the MaMMoS project and beyond for the following purposes.
\begin{itemize}
    \item Consistent user interfaces: The ontology will be used to define the input and output parameters of material simulations and characterisations in a consistent way.
    \item Creating interfaces for sharing data: These interfaces will facilitate the sharing of data in databases for magnetic materials, allowing data providers to use their native formats such as .csv, .yaml, or .hdf. The interfaces will store units and unique identifiers, allowing the quantities used in different datasets to be interpreted unambiguously.
    \item Developing data converters: These converters can automatically convert data between different formats and units, based on the definitions in the ontology. This would make it easier to share data between different research groups as well as between software and AI tools.
\end{itemize}

All information on how the ontology was built and how it can be used is openly available on the MaMMoS GitHub page\footnote{https://github.com/MaMMoS-project/MagneticMaterialsOntology}. There, all changes from within the project consortium will be published while other contributors can also discuss issues and suggest changes. Additionally, members of EMMC listed the ontology for MagneticMaterials as EMMO-related domain ontology on the GitHub repository of EMMO\footnote{https://emmo-repo.github.io/domain-magnetic-materials/magnetic-materials.html}.
The initial version of the ontology for magnetic materials was also published on Zenodo and can be found via \citep{fangohr_mammos-projectmagneticmaterialsontology_2024}.

\section{Conclusion}
This work presents a domain ontology for magnetic materials (MagMO) that addresses the semantic and structural challenges inherent in the field of magnetism. By formalising key magnetic properties across multiple scales and aligning with the Elementary Multiperspective Material Ontology (EMMO), the new ontology provides a consistent framework for describing intrinsic, hysteretic, and microstructural characteristics of magnetic materials. It supports semantic interoperability between modelling, characterisation, and data management tools, and facilitates the integration of magnetic material data into broader materials science infrastructures.
A central contribution of this ontology is its human-readable, programmatic implementation, which enables direct manipulation, versioning, and collaborative development without relying on opaque formats such as .owl or .ttl. This approach substantially lowers the barrier for domain experts to contribute and adapt the ontology to evolving research needs.
The ontology is openly available and designed for extension by the community\footnote{https://github.com/MaMMoS-project/MagneticMaterialsOntology} \footnote{https://doi.org/10.5281/zenodo.14547624}. This development, together with additional Python libraries such as mammos-entity that already use the ontology internally, supports FAIR principles and aims to improve traceability, reproducibility, and data exchange in the whole field of magnetism. Future work will focus on refining the structure of the ontology and the reuse of existing ontologies like CHAMEO, expanding its coverage to additional magnetic phenomena, fabrication, as well as characterisation technologies, and integrating it with emerging standards and platforms in materials informatics and data exchange.

\section*{Acknowledgments}
Funded by the European Union (Grant agreement ID: 101135546). The views and opinions expressed are, however, those of the author(s) only and do not necessarily reflect those of the European Union or the European Health and Digital Executive Agency (HADEA). Neither the European Union nor the granting authority can be held responsible for them.

\bibliographystyle{plain} % Not needed if specified in the document-class
\bibliography{references} % Entries are in the refs.bib file

\newpage

\end{document}